\documentclass[journal]{IEEEtran}
\usepackage[utf8]{inputenc}
\usepackage{graphicx}
\usepackage{amsmath}
\usepackage[square, numbers]{natbib} 
\usepackage{url}
\usepackage{xurl}
\usepackage{hyperref}
\usepackage{setspace}
\usepackage{titlesec}
\usepackage{float}
\usepackage{placeins}
\usepackage{adjustbox}

\providecommand{\keywords}[1]{
  \small	
  \textbf{\textit{Keywords---}} #1}
 
\titleformat{\section}[block]{\large\bfseries}{\thesection.}{1em}{}

\title{Vibe Coding Ate My Homework: An evaluation of AI approaches to greenfield software engineering and programming}
\author{Callum Barbour}

\begin{document}
\maketitle
\newpage
\begin{abstract}
Thanks to rapid developments in generative AI, we are in the midst of a paradigm shift that may change how we interact with computers forever. We have observed a growth in the use of natural language prompts to build applications and coding infrastructures without underlying knowledge of the field, and this practice has been dubbed `vibe coding.' It arguably represents what the field of programming has been building towards since the beginning, with every higher level of abstraction that is conceived. Vibe coding promises to be the endpoint for the meta of high-level programming as far as method of input is concerned: eliminating a human's use of code syntax entirely in favour of programming in their mother tongue. This paper aims to evaluate the viability of vibe coding for greenfield software engineering tasks, as well as analyse the benchmarks that have been used to measure its software engineering prowess. To this end, we have developed an evaluation suite for analysing an LLM's proficiency in carrying out simple, isolated greenfield programming tasks in Python to provide scoped insight on the matter.
\end{abstract}
\section{Introduction}
Automated generation of code has been a long time aim in computing \cite{nosilverbullet}, and while what qualifies as `automatic programming' has changed over time, it is ultimately the pursuit of programming at a higher level of abstraction than what is presently available to the user \cite{strategicdefense, noman}. Vibe coding ostensibly represents the natural endpoint of this development with natural language requests being the only input needed from the user while an AI agent manages everything else. The term vibe coding was coined by Andrej Karpathy, co-founder and former researcher at OpenAI \cite{karpathy}, describing the practice as `giving into the vibes' by letting AI fully control code generation without human review. While AI-assisted coding has existed for years \cite{aiassist} with human in-the-loop software engineers to review and fix any problems that the agent may have created, there has been a recent surge in programmers who do not review AI-generated code at all, in large part due to the perceived accessibility of vibe coding which has introduced an influx of new programmers without a coding background, such as those from clinical fields \cite{clinicalteaching}. This also includes corporations who have used the trend to justify layoffs, only to backpedal when they were not able to work around the shortcomings of AI \cite{layoffs}. Due to how quickly this practice has risen in popularity there is a pressing need to evaluate the reliability of AI agents in producing relevant and functioning code, as well as the benchmarks used to evaluate the execution of this task. With the barrier of entry to software engineering being lower than ever, a greater burden is placed on AI to produce reliable results, and its abrupt rise to prominence has been criticised as premature by some software engineers over concerns for code quality without human involvement \cite{visecoding, gotchas}. Potential applications of vibe coding cover all forms of programming tasks, and as with other methods of coding, all of them can be categorised as either greenfield or brownfield approaches \cite{greenfieldvsbrownfield} - the former refers to standalone programming tasks that are not tied to existing architecture. On a larger scale, they can refer to the development of an application or system from scratch. The latter concerns feature-level implementation tasks for a pre-existing system and are typically characterised as expansions or improvements to architecture already in place. The way prompts are expressed to an agent can reveal one's level of expertise in the field; higher level prompts are more broad and directly relay the needs of the user in a straightforward manner (e.g show me the numbers 1 to 10), whereas lower level prompts specify the means of achieving that goal in a manner that a computer can more easily understand (e.g create an array of numbers 1 to 10 and print each one in succession using a for loop). Those with a programming background have an inherently greater understanding of how computers think and carry out tasks which allows them to express their requests at a lower level of abstraction to improve the reliability of the output when needed as this reduces ambiguity \cite{betterprompts}. The general craft of writing better prompts is known as `Prompt Programming' and covers other techniques as well such as Chain-of-Thought (CoT) which allows the LLM to break down the prompt by asking it to `think' step by step before solving the problem \cite{promptprogramming}. Given the variation the prompt's ambiguity can have on the output, it is imperative that our evaluation of vibe coding's performance in greenfield tasks and the benchmarks used to measure said performance recognise the potential impact that greater knowledge in the field may have on output quality in order to properly ascertain the usefulness of vibe coding for both outsiders and seasoned programmers, which may allow us to pinpoint the minimum proficiency in Prompt Programming needed for vibe coding to be viable. With all of this in mind, we developed a rigorous greenfield evaluation suite that sends a dataset of prompts requesting Python programming tasks to four different LLMs to produce a vibe-coded output and determine if the task was successfully fulfilled based on that output. This has provided us with a strong proof of concept for potentially ranking contemporary AI models and software agents for end-to-end software construction with a transparent scoring protocol in the future. We created a list of test prompts and classified them into three different groups based on technical depth: 1. Surface level requests. 2. Requests for specific features and/or implementation expressed using vague or common language. 3. Requests for specific features and/or implementation applying low-level jargon. We believe this to be an important distinction compared to other evaluations of Large Language Models (LLMs) as they did not focus on the content of the query itself \cite{rigorousevaluation}. Finally, we chose to specialise in evaluating greenfield tasks and there are defensible reasons for this; LLMs are trained on a vast number of code repositories which all of their generated code may be derived from \cite{llmusage}. This introduces additional difficulties for brownfield tasks as they are required to fit a pre-existing system's workflow, and if the requirements are niche enough the model may struggle to produce a combination of code to fulfil the task without human input \cite{feabench}. While this is an avenue worth pursuing further in its own right, we found it to be less aligned with this project's chosen method of evaluating vibe coding as a long-term strategy (i.e because vibe coding is a hands-off approach to coding favoured by non-programmers, we believe its ability to generate isolated working programs from natural language prompts is ultimately what is relevant). Our hypothesis was that lower level queries would produce more reliable results at the cost of accessibility (which is the main selling point of vibe coding), a viewpoint that our final readings would challenge.

\section{Related Work}
Given the recent development of vibe coding, its benchmarking is still in its infancy. Not only has it merely been explored by a few select papers, but the work that has been done has been criticised as not measuring the right things as well as half of the benchmarks claiming to measure abstract ideas like reasoning with no clear definition of what that means or how it is measured \cite{badjoke}. One study \cite{featbench} suggested that existing evaluation benchmarks do not adequately assess an agent's vibe coding capabilities. To that end, they proposed a novel benchmark named FeatBench with a focus on feature implementation. This study also noted that issue-solving benchmarks as represented in \cite{swebench} are most relevant to the capabilities of vibe coding, as they build code based on issue descriptions, but they tend to neglect other aspects of its vibe coding performance such as feature implementation. Ultimately, they determined that AI agents tend to adopt an aggressive implementation approach. Without careful specification, AI agents can go far beyond the intended scope and create errors that may be difficult to troubleshoot without manually inspecting the code. However, this behaviour does come with the benefit of potentially creating superior and more robust software architecture than envisioned. This highlights a need for a way to control the implementation aggressiveness of a vibe coding agent. A notable limitation of FeatBench is its exclusive focus on Python for evaluation, which does not adequately cover the scope of a greenfield project. Our initial plan to expand upon this was to represent one language for three different `genres' (Markup, Scripting, Object-Oriented) which would be HTML, JavaScript, and Python. Unfortunately, due to a combination of time and resource constraints the system had to be scaled back to just Python. However, our working system is built in R, which serves as an effective hub to bridge multiple languages together, should an expansion be greenlit. Additionally, where FeatBench focused on feature-level implementation (i.e brownfield programming), our system measures the models' capabilities in performing simple tasks that don't revolve around an existing system. A separate paper \cite{emergingtrends} established a three-pillar framework for guiding responsible adoption of vibe coding based on the results of a performance evaluation - hybrid integration, human oversight, and context-aware deployment. They evaluated the performance of traditional coding without AI, human-in-the-loop AI assistants and fully autonomous vibe coding agents and concluded that vibe coding should not replace AI-assisted coding, that human developers maintain responsibility for validating outputs and that the practice should be limited to non-critical applications, education contexts and rapid prototyping until the current quality and security concerns are addressed. The results of said evaluation revealed that vibe coding can improve efficiency by up to 27\% at the expense of maintainability, security and user trust, sentiments supported by \cite{itleaders}. Overall, it obtained a System Usability Scale (SUS) score of 71.4, placing it above traditional coding but below AI-assisted coding. Developers were surveyed for qualitative feedback, and the results highlight the mixed reception of the user experience, with 63\% reporting less mental strain related to syntax, while only 37\% displaying full trust in vibe coding outputs without manual review. This study provided a solid foundation for our own evaluation suite, but we adapted it to measure the proficiency of greenfield tasks, as well as differentiate its reliability and quality of output based on the level of expertise expressed by the query. Maintainability concerns have been echoed by \cite{maintain}, proposing a framework which recommended a rigorous degree of human oversight - however, the inherent flaw is that what sets apart the term `vibe coding' from other forms of AI-assisted coding is that human engagement with the code is kept to a minimum \cite{futureoutlook}, if not cut out entirely. As such, the delegation of maintainability to a human prevents vibe coding from being entirely accessible in projects where such matters are a concern. Still, the development of an agent that can apply newer coding conventions with proper prompting may be a potential automated alternative in the future. Currently, maintainability is not something that can easily be measured or evaluated by a test suite as it is a continuum that is both partially subjective (ease of human understanding may vary from person to person) as well as sensitive to future developments and versions. Security has also become a major problem among large companies who have introduced vibe coding into their discipline. This was extensively covered by YouTube user Logically Answered \cite{logicallyanswered}, who reinforced the reality behind the functionality of LLMs, being that they are not truly logical and can be summed up as an algorithm that predicts the most likely next word or action and creates it. They point out that consequently AI models do not truly know if the code they are producing is correct, rather they are producing code they deem most probable based on the public code they are trained on, and since public code is often subpar, it has given rise to very inefficient code with deep architectural flaws and vulnerabilities to the point that some programmers prefer starting from scratch themselves instead of fixing all of the problems \cite{dougdoug}. Itay Nussbaum of Apiiro noted that while syntax errors were reduced, architectural flaws such as privilege escalation ballooned by 322\% \cite{timebombs}, inspiring his remark that `AI is fixing the typos but creating the timebombs.' For this project, we settled on examining the general reliability and functionality of vibe code for the purpose of establishing a strong and expandable foundation. However, metrics such as security could become a viable benchmark for future iterations of this system designed to evaluate more complex tasks. LLMs being functionally prediction machines can be considered a core cause of the common hallucinations that arise; an honest AI-agent that replies `I don't know' to a request they don't have the means to carry out guarantees an `incorrect' answer whereas putative guessing has a chance to produce the correct response regardless of how slim its odds are. As such, an LLM which hallucinates due to lack of sufficient training will appear to perform proportionally better than an honest agent over a large scale of tests. This hypothesis is supported by a study \cite{humannatureghost} which determined that AI has a habit of prioritising appearance of competence over admitting its limitations, resulting in systematic deception patterns. Combining these findings with what we know about LLMs' probabilistic nature, we can deduce that the resulting behaviour is not exclusively about maximising the probability of correctness, but rather maximising the probability of what it believes to be `the ideal answer,' in which the manner of expression becomes relevant. We can theorise its thought process as two steps: 1. Find the most probable solution available. 2. Assert it confidently and authoritatively. Granted, one could argue that step 2 is simply a consequence of AI predicting the most likely next word and thus retains a common objective tone, but the observed outcome is nonetheless the same, even across different agents as shown in the aforementioned study. An LLM may therefore interpret a correct response expressed confidently as closer to the `ideal answer' than a correct response suggested cautiously, and while this is true on paper, in practice it would mean that an agent will present their answers with convincing and seemingly informed language by default regardless of their exactitude and this has the consequence of potentially misleading humans if their code is not properly reviewed. This is an unfortunate consequence of probabilistic machines - by prioritising the probability of `the ideal answer', they are incapable of admitting their limitations or displaying uncertainty as either of these actions prevent that event entirely. With these limitations in mind, we get a clearer picture of AI's potential trappings in greenfield tasks and how an experienced programmer might work around them, and this shall be reflected in our test prompts - lower level queries will incorporate practices that split a greenfield task into easily automated and widely documented actions where possible.

\section{Methodology and Technical Approach}
\subsection{Scope}
As specified earlier in this article, this project is mostly concerned with measuring the reliability of LLMs in carrying out simple greenfield tasks. We decided to evaluate four locally run Ollama models on an HPC cluster. Due to the conceptual complexity of such a system, what our system represents is a small-scale MVP meant to represent a working proof of concept. The system is currently only designed to support Python. While most of our scoring logic is language-agnostic in principle (particularly artifact-heavy cases such as for images and spreadsheets), new pipelines would need to be developed during the prompt creation and syntax check stages, and this branch could be implemented as an additional loop within run\_evaluation with a conditional to distinguish the language each task belongs to. We would also require the development of further wrapper templates to accommodate each new language which could most easily be accomplished by expanding the wrap\_code function to assign different templates based on the language associated with the task. In this project, we identify vibe coding as a hands-off form of programming in which the user can create programs through natural language descriptions and does not directly engage with the underlying code at any point. Through this interpretation, it becomes clear that the scoring system must be absolute. An implementation that is mostly correct, but contains an error that crashes the whole program, for example, cannot be given any credit as the user cannot be expected to manually fix or understand what is wrong as that falls outside the practice of vibe coding. In addition, we included a retry system that triggers in the event of a runtime error. This design choice was inspired by the fairly common use case where the user will ask the AI to fix the code in the event of it failing to execute, providing available execution evidence such as captured console output, observed variables, failure status as well as any external execution output where available. It allows us to assess an LLM's capability of course-correcting after a hallucination, while also distinguishing them from models that managed to execute the first time without fail.

\subsection{Evaluation Suite Architecture}
R is the system orchestrator and is the root from which the entire functionality of the system is derived. It starts by fetching the task text files inside each of the level directories, recording the level directory it was found in, the name of the task file, and the contents of the text file (the task prompt) to be mapped to a list, which is then added as an entry to a global list; this is done for every task file. After this, the tasks (now represented as list objects in R memory) are checked to find the lowest level prompt of a given task ID (for our system this will be level1 for all samples) and this is recorded as an added field called the scoring\_prompt. The reason behind this was to ensure the metrics for passing and failing remain fair across every level, as more technical and detailed prompts could in turn be judged more harshly by the scorer due to the greater specification in the prompt it uses to judge the output. By using the most surface-level iteration of the prompt for every level of a given task ID, we ensure an equal scoring standard across all of them. After the scoring prompts are assigned, the evaluation phase begins. First R sends the system prompt concatenated with the task prompt for that cycle to the model under evaluation using a fresh interaction session for each iteration. The system prompt contains instructions for structured output via wrapping the code in JSON with a \texttt{code} field. After the response is retrieved by R, it must then parse the response in order to extract and validate the syntax of the generated Python. After this, R creates a workspace folder (if it does not already exist), in which to store more folders pertaining to the output of each iteration of the evaluation. R makes use of a pre-built code wrapper intended to be used for a .py script to denote system tasks that must be pre-programmed on Python's end. These relate to the construction of the result.json file and the retrieval of the data used to build it, comprising the global variables (or first function-level if the whole code is wrapped in a function), the console output observed during execution, as well as whether or not the program successfully executed. The LLM-produced code is assigned to a variable in the wrapper which is then fed to an exec function for the purposes of executing the code within a tracking namespace that allows for the observing and recording of variables by both name and value. After the wrapper is written to a .py file, the script then begins execution. The space in which the LLM code is executed is wrapped in a try block so that in the event of an execution failure the entire script will not halt and instead record the failure state in the result file. In the event of a failure, the LLM is given one chance to correct their mistake with a new implementation, sharing some observed data throughout the previous run. Upon a successful run, R checks the output produced by the LLM and funnels it through one of three pipelines based on the presence of certain types of files. If an Excel file is found, the contents are inspected computationally without the aid of a model. If an image file is found, the contents of the image are examined by a vision model to decide if it met the task requirements. In all other scenarios, the list of variables and console output stored in the result file is checked by a text model instead. In the latter two scenarios, the scorer models are sent a prompt instructing the required response format, as well as the lowest-level task prompt associated with the ID of the task that was prescribed to the LLM under evaluation. It then responds with a JSON object containing a `passed' field that is either true or false based on the LLM's judgement. Concluding each evaluation of a sample, the results are represented as a list containing background information such as the model under evaluation, the ID of the task it is being judged on as well as the level to which the task belongs. The relevant scoring metrics captured on the list cover the syntax validity on the first attempt, a functionality score based on how many attempts it took the LLM to produce error-free code, as well as whether or not it successfully fulfilled the task. Additionally, an \texttt{output\_reason} field exists to provide further information behind the motivation for the verdict if it failed, while simply reading `correct' if it managed to pass. These results are then appended to a list for every iteration of the evaluation process, and when evaluation concludes the results are saved to both an .rds and .json file. The pipeline of the architecture is showcased in Figure~\ref{fig:pipeline} below.

\begin{figure}[H]
    \centering
    \includegraphics[width=\linewidth,height=0.7\textheight,keepaspectratio]{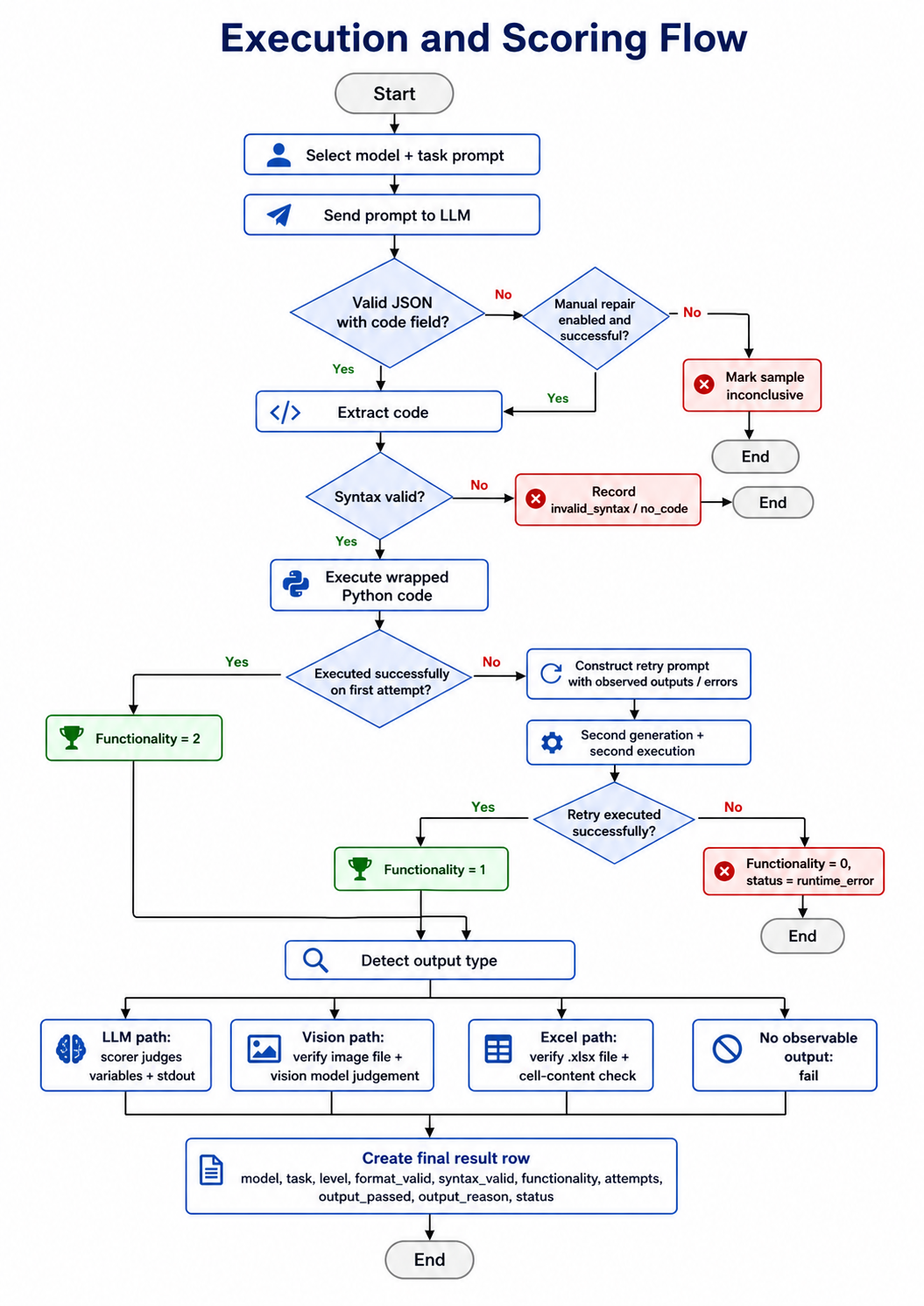}
    \caption{Evaluation pipeline for the greenfield benchmarking system.}
    \label{fig:pipeline}
\end{figure}

\subsection{Scoring Approach}
\label{sec:scoring-approach}
\begin{figure}[H]
    \centering
    \includegraphics[width=\linewidth,height=0.7\textheight,keepaspectratio]{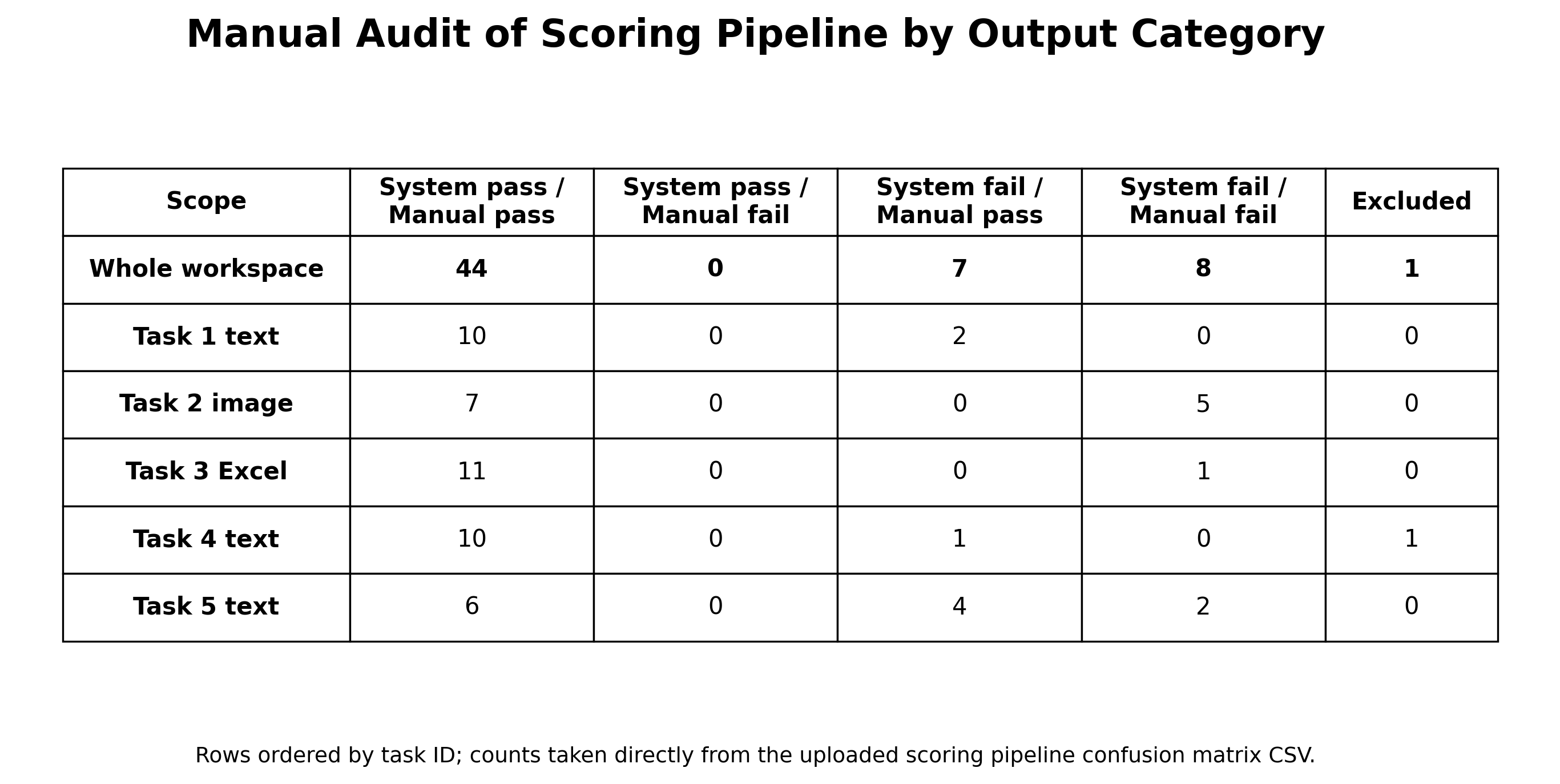}
    \caption{Manual audit results.}
    \label{fig:cm}
\end{figure}

The metrics we have used for assessing the code's adequacy remained consistent for every task that was given. However, the forms of output each task is expected to take are too diverse for them to be processed the same way. Thus, a text model is used to assess purely computational tasks (i.e no creating or writing of external files), a vision model is used for cases where an image file is produced, and finally a deterministic inspection of cell contents when an Excel file is produced. Our motivation for using extra LLMs for scoring most tasks was to account for the nebulous range in which a task's output can be expressed while sharing the same semantic substance \cite{judge}. Task 1, for example, involves the creation of a randomly generated array and extracting the highest number contained in it. Semantically, verifying the output's adherence to the task is simple - observe if an array exists, and whether or not a variable holding the maximum value is observed. Syntactically, however - it is infeasible to comprehensively represent the ambiguous spectrum that a correct answer could take based on the unpredictable variation in names that an LLM could choose to represent the maximum number. Furthermore, a variable's objective purpose is an interpretive matter which cannot be expressed as a computational problem. One could argue that simply recording the value would sufficiently accomplish the task, but as we are evaluating the model's adequacy as a vibe coding agent, we feel that semantic clarity is an important factor and that the name attributed to the variable must not misrepresent its value. As a compromise, we resorted to utilising the probabilistic nature of LLMs to provide a practical method of parsing and analysing more ambiguous tasks like these. However, the absence of a deterministic, hard-coded scoring process for tasks that don't involve writing values to Excel spreadsheets introduces a risk for false readings to occur. We felt it prudent to perform a manual audit of the judgements each scorer provided and compiled confusion matrices for each task as well as one indicating the overall reliability of our scoring pipeline (Figure~\ref{fig:cm}). The excluded field denotes inconclusive cases (i.e a failure to evaluate the code, typically due to broken JSON). Only one such case took place in our final run of the program.

These readings showcase several false negatives, particularly concentrated in task 5 while there are no false positives. This suggests that our scoring model is somewhat conservative. It is possible these false negatives were derived from the scorer being uncertain how to handle irrelevant data that may leak into the result.json's variable list. One of the false negatives concerned Gemma's level 2 task 5 run. Even though the task had been successfully accomplished upon a manual inspection, the memory address for a list iterator object had leaked into the variable list and this bloat may have influenced the scorer negatively. This explanation is substantiated by the fact that every text-based task contains at least one false negative among its runs, while the Excel and image-based tasks - where scoring is judged based on the existence of a file as opposed to the information on the result.json file - do not contain any false negatives. However, it should be noted that our image scoring model is currently hard-coded to handle task 2 specifically and is not built to detect images that are not straight line graphs. Nonetheless, the vision model's consistent ability to correctly identify straight line graphs is notable since each test sample is conducted under a fresh session to ensure fair comparisons across each task and level. The Excel method did not rely on a model for scoring and was carried out via deterministic computation, which eliminates the risk of false readings, provided the process correctly represents the task requirements. Like the image scoring path, it is designed to accommodate task 3 specifically and will require future modification to achieve general functionality.

\subsection{Experimental conditions}
\subsubsection{Controlled/independent variables}
Given the number of dimensions in this experiment, the line between controlled variables and independent variables is somewhat blurry as it largely varies on which metric the user chooses to view everything in relation to. However, at the highest level our closest analogue to a controlled variable would be our system prompt specified before the task instructions for every sample tested. They are system level instructions responsible for ensuring structured output that can easily be parsed and transferred across different programming languages by ordering the LLM to produce a response that is itself a valid JSON object containing a field labelled \texttt{code} in which to store its code fulfilling the task. Similar equivalents exist for model-based scoring, as well as retry attempts, however their applications are conditional as opposed to the main system prompt that is called at the beginning of every evaluation instance. This is because each call to an LLM is performed in an isolated session with no prior memory, and thus these parameters must be specified alongside the sample-specific attributes. As alluded to previously, the nested structure of progression gives the models, tasks, and levels flexible classifications in terms of variable groups at the local level (globally they are all independent variables) according to what trend the user is interested in focusing on. Since this project centralises around evaluating the performance of models, model identity is a primary independent variable in the overall experiment - however, under the interpretation of each model's evaluation being an isolated experiment, it is kept constant. The performance of each model is evaluated based on its execution of five unique tasks at three different levels of technical detail - making for a total of 15 prompts they are individually required to produce results for (our system used four models for evaluation, giving us a total of 60 data samples). Task prompts mark a less formulaic and more broadly semantic form of data in our project, as their contents correspond to plain English text. As such, there are unfortunately no rigidly systematic distinctions between tasks across different levels and can be somewhat subjective. Regardless, we have attempted to properly categorise and standardise them by having them follow a general template for their expression, with level 1 tasks being categorised as purely surface level, conversational requests, level 2 providing more descriptive functionality and finally level 3 doing the same with reference to the proper computer terms to minimise ambiguity. Matching task IDs across different level groups are expected to accomplish the same goals based on what is provided in the info.txt file attached to the project. The system currently has no method of enforcing this rule, and its adherence must be ensured manually. The varying level groups (level 1, level 2, level 3) represent the different degrees of technical expertise demonstrated by a prompt. Vibe coding may primarily be practiced by non-programmers, but in the interest of a more conclusive assessment we felt it was important to acknowledge how one's familiarity in the field may impact the reliability of such a practice. Thus, matching task IDs in different level groups represent different methods of expressing the same task.

\subsubsection{Dependent variables}
The dependent variables for this study can be broadly defined as the LLM's adherence to the task. However, this concept is represented through several different metrics - namely syntax validity, functionality score, and an overall passing verdict. The syntax validity reading is determined after R compiles the generated Python code extracted from the JSON object. If the code fails to compile then the sample will be flagged as having invalid syntax. The sample is still permitted to proceed to the execution stage as normal and will trigger the retry process upon its inevitable failure. The invalid syntax reading is maintained even after the code successfully executes on the second attempt as it is never updated after the first check. This is to give us further insight on what caused the code to fail initially. On the topic of retries, this is where the functionality score becomes relevant. This parameter more so concerns general stability as opposed to correctness, but if the code fails to execute for any reason then R will send another prompt to the LLM telling them that their previous attempt failed to produce results, sharing relevant information such as any variables or console output leading up to the crash. We intended to capture the error message too, but did not realise we were not capturing it properly until after the development stage concluded. In addition, the functionality score will be decremented to 1. If the LLM fails the second attempt, the score will be reduced to zero and the task deemed a failure. The passing verdict is self-explanatory and is represented as a boolean field - true readings mean the model passed, and false readings indicate it failed. In addition, there is a field called '\texttt{output\_reason}' to provide further context for the LLM's decision (e.g whether it failed because of an error or simply incorrect output).

\section{Results and Analysis}
The results from this experiment are showcased in Table~\ref{tab:model-performance-comparison}. There are two sets of readings - the first represents the pass rates for the results as produced by the suite. The second set showcases the genuine performance rates after a manual audit was performed to eliminate false judgments. As previously suggested in Section~\ref{sec:scoring-approach}, these results indicate a strict scorer. In terms of the model performance, we see that Phi4 performs notably better than the others on both counts with Gemma and Mistral being roughly on par with one another. However, Qwen\_coder has by far the biggest discrepancy between its predicted and actual performance. It is possible that Qwen\_coder's methods of generating the required code are less compatible with the tracking namespace we have set up, and consequently led to a higher risk of irrelevant data leaking into the result.json file, as was the case with its level 3 task 1 run. While this logically does not stop the output from being correct, it is possible that the noise introduced by bloat such as a list\_iterator object memory address may have confused the scorer, leading to a false negative.

\begin{table}[H]
\centering
\caption{Raw and manual-adjusted model performance.}
\label{tab:model-performance-comparison}
\begin{adjustbox}{max width=\linewidth}
\begin{tabular}{lrrrrr}
\hline
\textbf{Model} & \textbf{Raw passes} & \textbf{Raw rate} & \textbf{False negatives} & \textbf{Adjusted passes} & \textbf{Adjusted rate} \\
\hline
\texttt{phi4}        & 13/15 & 86.7\% & 0 & 13/14 & 92.9\% \\
\texttt{gemma}       & 11/15 & 73.3\% & 2 & 13/15 & 86.7\% \\
\texttt{mistral}     & 11/15 & 73.3\% & 1 & 12/15 & 80.0\% \\
\texttt{qwen\_coder} & 9/15  & 60.0\% & 4 & 13/15 & 86.7\% \\
\hline
\end{tabular}
\end{adjustbox}
\end{table}

Once we look at the pass rates relative to the prompt levels (Table~\ref{tab:prompt-level-performance}), we see an interesting trend that does not line up with our hypothesis. Our assumption was that greater technical expression would improve the consistency of a model's performance, however the results of this run appear to suggest a slight decline in genuine performance, and a sharp 30\% drop as judged by the scorer. This indicates that whatever minor problems are to blame for this drop-off are likely more pronounced when taking into account the scorer model's strictness. We believe that the problem may be model-bound, and that the longer and more detailed task prompts may be overwhelming the models we have chosen to evaluate, especially when they are bundled with a system prompt that relays even more instructions on top of that. It is also possible that despite the expectation more elaborate instructions would reduce ambiguity, it also creates more variables the LLM needs to consider and thus more variables it could potentially misinterpret. One notable example of a semantic misunderstanding is found in Mistral's level 2 task 5 run. Rather than producing a random array to then rearrange alphabetically, it instead requested the user to input the string to be rearranged. The task prompt this output stemmed from began with `Given a string made up of 12 random letters...' which could suggest that the model did not interpret `random' to mean the random package, but rather a random string provided by the user. This is likely due to the use of the word `given' which, given its semantic meaning, may have led to the model interpreting random in a different context due to its probabilistic nature. This would suggest that words with less flexible meanings should be favoured in the interest of more consistent output. Given the highly simple programming tasks we used for this project, it is reasonable to conclude that the advantages more technically detailed language affords are better suited to more complicated tasks and stronger models, and do not outweigh the drawbacks under the conditions used for this study.

\begin{table}[H]
\centering
\caption{Raw and manual-adjusted pass rates by prompt level.}
\label{tab:prompt-level-performance}
\begin{adjustbox}{max width=\linewidth}
\begin{tabular}{lrrrrr}
\hline
\textbf{Prompt level} & \textbf{Raw passes} & \textbf{Raw rate} & \textbf{False negatives} & \textbf{Adjusted passes} & \textbf{Adjusted rate} \\
\hline
Level 1 & 17/20 & 85.0\% & 1 & 18/20 & 90.0\% \\
Level 2 & 16/20 & 80.0\% & 1 & 17/19 & 89.5\% \\
Level 3 & 11/20 & 55.0\% & 5 & 16/20 & 80.0\% \\
\hline
\end{tabular}
\end{adjustbox}
\end{table}

\section{Limitations}
Figuring out the limitations this concept had in execution and working around them was a large part of the development process. We wanted to assess state-of-the-art (SOTA) models such as ChatGPT and Gemini for a more up-to-date analysis on the state of the industry. However, a credit system is in place for most of them which require a fee for tokens used. In order to avoid financial expenditures during the debugging process we decided to assess locally run models hosted by Ollama. Given the limited disk quota each user is given on the cluster, we could not fit the most advanced models on our working environment, and so we compromised by selecting the four strongest models that were able to fit on our cluster for evaluation, as well as two scoring models, one for text-based tasks and the other for image checking. Another limitation of our program was our limited sample of tasks - five tasks expressed at three different levels of technical depth, for each of the four models, giving us a total of only 60 result samples. Given how resource-intensive this suite is even with HPC, the time taken for the evaluation suite to conclude was roughly 20 minutes, and this was the main reason we decided to limit the number of unique tasks represented. To make up for this, we made sure there was sufficient variety between each task based on the form of their expected output (e.g image creation, raw computation etc.) as it enables us to have a more feature-complete and versatile proof-of-concept. The scoring function R decides to use for evaluating a task is not technically based on the type of task performed, but the observed output. If an image task failed to produce an image, for example, R would analyse the output via the text model instead. Given the dataset we are working with, this fortunately does not cause any serious issues. A common theme for many limitations in this project would be `ambiguity.' Our samples are conversational natural language prompts that cannot be objectively standardised. Furthermore, while we provided clear guidelines on what form the prompts take in each level, there is still a degree of subjectivity involved when deciding what qualifies. We additionally decided to rely on LLMs for scoring tasks 1, 2, 4 and 5 for a more general purpose-friendly pipeline (although the vision path is still partially hard-coded for task 2), as this allowed us to avoid accounting for every possible form a correct answer could take. However, it also comes with a risk of false readings due to hallucinations deriving from its probabilistic logic. Currently, this system only supports Python. R was selected to serve as the central hub for a suite that could analyse several different languages. As development progressed, we felt it was more practical to scale back functionality to exclusively Python, allowing us to focus on building a raw MVP that can be further built upon. One feature of our suite in particular that demands more work would be our implementation of the retry process. In failure cases that produce errors, R is supposed to capture the traceback containing the error message to provide extra context to the LLM on what went wrong before commencing its repeated attempt. Unfortunately, while the Python wrapper is built to catch exceptions internally to prevent the halting of the entire Python program which is responsible for providing important execution details for evaluation, we were under the misunderstanding that details of the error message would be stored in stdout alongside standard console output, and so our retry prompt was constructed under this assumption. In reality, traceback is typically stored in stderr instead, but since our wrapper catches those exceptions internally, they are not emitted at all. As such, the only information the LLM has to work with is that its code failed to execute - along with any notable data created before its crash such as variables and console output. This should be prioritised as a fix in future revisions due to being an unambiguous fault in our system's functionality that was not discovered soon enough. As of right now, the biggest question regarding the limitations of our system is whether or not it truly represents the current capabilities of vibe coding. The purpose of this project was to establish a greenfield evaluation suite for analysing LLM generated code. What has been achieved is a working proof of concept that can be further refined to feature more languages to use for evaluation, more task prompts to include, as well as the hosting of far stronger models. The answer to the aforementioned question, however, is that vibe coding is a spectrum that can vary from simple programming tasks, all the way to large applications and coding infrastructures, and so what we can conclude is that our system represents vibe coding's capabilities at the most basic level - however, real-world applications of vibe coding often involve the development of entire software systems and codebases. With this comes a significantly wider range of parameters for assessing vibe coding's merit, such as security, or algorithm efficiency - two metrics that are not represented in any form in this suite. Finally, it should be noted that while we decided that every prompt sent to an LLM, whether it was a scorer or an examinee, be sent in a fresh session with no memory retainment for the sake of fairness, this does ignore a critical aspect of LLMs, as while they do not update their trained parameters over the course of a session, they do demonstrate the capability to adapt their responses through the conversational context \cite{prefeval}, and this may be a useful benchmark to consider for evaluating LLMs in the context of code generation. This restriction may also harm our ability to properly assess a model's potential to fix an incorrect output, as our retry system is still set up to communicate through fresh sessions and so much of the context used to carry out the task must be repeated back to the LLM to compensate for this, but this does not change the fact that the LLM has no genuine recollection of producing the code it has been told to fix and must attempt the task from scratch, rather than perusing what it has already generated to eliminate any syntactic or semantic flaws.

\section{Discussion}
Based on our findings, we can conclude that LLM-generated code is mostly reliable, albeit somewhat volatile. However, if problems can already occur at a feasible rate when the tasks are kept this simple, then it calls into question the longevity of vibe coding when the tasks are made more demanding, as they frequently are in real-world use. If simple programming tasks such as these invite a reasonable risk of incorrect output (as was demonstrably proven in our results), then requesting the creation of larger software infrastructures, composed of many comparably simple subtasks may compound these opportunities for error, particularly without human review or debugging which would betray the practice of vibe coding. In these circumstances a vibe coder is reliant on the model to fix the errors, whose suggested fixes could introduce further errors. For these reasons, we can assume that an entirely hands-off approach to programming (otherwise known as vibe coding) is not viable long term when it comes to larger, higher stakes software development, at least for the time being. Our hypothesis that more technical prompts would ensure more reliable output did not hold under the conditions of this suite. However, our findings gave us further insight into what constitutes a quality prompt. Using words with less flexible meanings (e.g. avoiding words like `given') can minimise unintentional behaviour. Furthermore, too much detail for a comparatively simple task can lead to diminishing returns as well as introduce higher hallucination risks due to the greater number of parameters that the LLM must probabilistically interpret which also means a greater number of opportunities for a false reading to occur. With this in mind, we can conclude that requests should be as concise relative to their scope as they are informative to maximise consistency of results. Phi4 statistically is the best-performing model in this suite, with Qwen\_coder being the most misrepresented by the notably strict text scoring model. We previously suggested that the possible cause could be due to noisy data making it into the result.json file. This would indicate that the text scorer's conservative nature stems from its rejection of superfluous data, and hence the logic dictating its decision is not based on whether the task was fulfilled, but rather if ONLY the stated instructions were carried out (interpreting the noise as an extra instruction). In most cases these two scenarios go hand in hand, but it is nonetheless a notable distinction that would explain the text scorer's strictness. The manual audit revision revealed that the true performance of each model was roughly on par with one another. Even so, the text scorer's false negatives do provide useful insight into how each model structures their coding output, and how some methods are not as seamlessly compatible with the tracking namespace in which the code is executed.

\section{Conclusion}
We have built an evaluation suite for models based on their proficiency in carrying out simple greenfield programming tasks in Python. We found that there are downsides to consider when opting for more in-depth requests and that striking a balance between brevity and context is more important for minimising hallucinations. Our goal going into this study was to present a reproducible harness and dataset that could be used to rank contemporary models and agents for end-to-end software construction. While we did not develop a system that definitively achieves this, we successfully built a scoped reproducible MVP comparing locally deployable models via their responses to greenfield Python tasks, and the results we have obtained do provide insight into answering that question. For a hands-off coding approach to software construction to be viable, the models must have mastery over simple programming tasks as software architecture is often composed of many such tasks. The adjusted accuracy of the models we tested falls within the 80-90\% range. This suggests that they are adequate for carrying out requests for specific features. However, relying on a goal-centric request to build an entire system is highly likely to introduce faults based on these results as it is composed of many individual programming tasks and features that the LLM is not even specifically prompted for and chooses based on a probabilistic process, and this compounds the opportunities for failure. For instance, a failure rate of roughly 1 in 10 per task becomes much higher in the context of a full codebase that contains far more than 10 individual sub-tasks. This means that error handling will be a necessity for AI models in vibe coding sessions. With no human intervention, it is left to the AI to figure out how to fix bugs they themselves have produced.

\section{Future Work}
There are many directions in which the system developed thus far could be taken. One possible extension would be the introduction of a larger task set that showcases greater variety in the forms of output observed. Currently, our set of tasks only captures a limited range of the applications of greenfield programming. By extension, our system is only built to parse the forms of output covered by our 5 main tasks. If future work aimed to develop a single universally applicable scoring pipeline, an immediate solution could involve the introduction of a stronger general-purpose model should the resources become available. One could argue that using LLMs as assistance for scoring could weaken the objectivity of our findings due to us effectively using the same tools for evaluation as the tools we are trying to learn more about. However, we feel it is important to acknowledge the versatility LLMs have as tools - and that their capabilities of reading and analysing results does not overlap with their potency in generating correct and functional code. Prior work \cite{judge} suggests that the probabilistic nature of LLMs lends them credibility as flexible evaluators for outputs where correctness is verified semantically and cannot be reduced to a strict algorithm. Therefore, we believe that there is merit in utilising LLMs for automating evaluation in cases where output cannot be properly evaluated via rigid deterministic metrics and the system could potentially be streamlined into only requiring one general-purpose pipeline for scoring. We should not overlook opportunities to apply deterministic scoring methods where possible, as the capability of verifying a task outcome deterministically can eliminate the risk of hallucinations introduced by probabilistic LLMs. As our system is currently built around only handling isolated and basic programming tasks, various mechanisms would need to be expanded to accommodate more ambitious queries such as the complete development of a software application. This would also open the door for further metrics we could measure, such as the maintainability, security and algorithm efficiency of the code - giving us more comprehensive methods of both analysing and comparing the performance of LLMs. By expanding the scope of our tasks, the necessity of using an external LLM for scoring becomes greater. This is because it is easier to develop hard-coded benchmarks for fundamental tasks whose conditions for correct implementation cannot be broken down into smaller tasks, though even this can present its own challenges. Task 1, for instance, concerns the development of code that creates a randomly generated array of numbers, and then finds the highest number from said array. Here, it would seemingly be easy to develop a rigid algorithm for such a task as there are not any ambiguous smaller tasks that you could further break this objective down to beyond what is already provided. However, even in a small task like this a conundrum presents itself - given that we are studying the use of vibe coding specifically, necessitating the use of more conversational and vague language representative of its majority demographic, how do we decide what constitutes a satisfactory method of `finding' the highest number? Should code print the answer to give the user a visible reference to it, or is the mere calculation sufficient? Under normal circumstances we would be more specific about our desired result, but we risk over-fitting our test prompts with superfluous specification that could make them less representative of how a vibe coder allocates work to a model in the real world. If more elaborate and measurable metrics were developed for this system, as well as allowing the system to run on a platform with access to more powerful GPU resources, increasing the number of runs per prompt courtesy of the \texttt{N\_RUNS} configuration could produce more fruitful information regarding a model's performance on the task based on averages in numerous different categories (security, correctness, maintainability, efficiency etc.).

\bibliography{references}

\section{Appendix}
The early stages of the project were spent gathering as many sources as possible and combining their findings with each other to formulate new theories as well as uncover new avenues for research. We used a search string on Google Scholar to gather a large initial pool of papers, most of which directly reference vibe coding, and we logged each potentially valuable paper in a google spreadsheet, which can be accessed \href{https://docs.google.com/spreadsheets/d/1NmDr89SAiwT4MTixhlrd-DAISePgY5KZci99Er7SRG4/edit?gid=0#gid=0}{here}. Our goal was to provide a solid foundation for addressing the research questions, which is why each source had columns for extracted information that potentially related to or answered these questions. Additionally, we decided to apply the same search string for Google's regular search engine for grey literature in the interest of holistically capturing the current zeitgeist of the industry. As this report developed, we began including other sources outside that pool for reasons such as backing up a claim, providing background for our line of thinking, or through snowballing techniques \cite{mlr}. This allowed us to identify new leads by comparing findings of different sources and establishing links between them to deepen our understanding of the field, in particular our proposal that the probabilistic nature of LLMs is the direct cause of its hallucinations in the face of ambiguity. Two sources, namely \cite{nosilverbullet} and \cite{badjoke}, were suggested by this project's supervisor while discussing feedback and research questions. In addition, a few sources were found with the assistance of ChatGPT when we wanted papers with specific information in mind (usually evidence supporting a claim). Our prompts simply requested articles and blogs that reference this information in some way. We felt it prudent to prioritise thorough research over developing preliminary code at that juncture in order to properly grasp how our software systems should be built, and if they are required at all. This eliminates any time-sinks that may arise from necessary overhauls and sunk cost fallacies. Our research so far has revealed gaps in our knowledge of AI approaches to greenfield programming, necessitating the creation of our proposed evaluation suite. Once our development suite began, all of our time was dedicated to building the system with liberal assistance from a variety of coding agents, most predominantly ChatGPT 5.4 and Claude. We conducted weekly meetings with the project supervisor in the interest of tracking progress, as well as recommendations for project direction.

All code, data variables and results for the run represented in this thesis are stored in \href{https://github.com/CallumBarbour/Vibe-Coding-Ate-my-Homework}{this GitHub repository}.

\end{document}